\documentclass[pra,twocolumn,preprintnumbers,amsmath,amssymb,superscriptaddress]{revtex4}

\usepackage{url}
\usepackage{amsmath,amssymb}
\usepackage{mathrsfs}
\usepackage{ifthen}
\usepackage{ifpdf}

\ifpdf
\usepackage[pdftex]{graphicx,hyperref}
\else
\usepackage[dvips]{graphicx}
\fi

\newcommand\ket[1]{{ |{#1} \rangle }}

\newcommand{\ignore}[1]{}

\newcommand{\capt}[0]{(Color online) }

\newtheorem{theorem}{Theorem}
\newtheorem{proposition}[theorem]{Proposition}

\newtheorem{lemma}[theorem]{Lemma}

\newtheorem{corollary}[theorem]{Corollary}

\newcommand{\qedsymb}{\hfill{\rule{2mm}{2mm}}}
\newenvironment{proof}[1][]{\begin{trivlist}
\item[\hspace{\labelsep}{\bf\noindent Proof#1:\/}] }{\qedsymb\end{trivlist}}

\begin{document}
\title{New lower bounds on the non-zero capacity of Pauli Channels}

\author{Jesse Fern}
\email{jesse@math.berkeley.edu}
\affiliation{Department of Mathematics, University of California, Berkeley, California, 94720}
\affiliation{Berkeley Quantum Information Center, University of California, Berkeley, California 94720}

\author{K. Birgitta Whaley}
\affiliation{Department of Chemistry, University of California, Berkeley, California, 94720}
\affiliation{Berkeley Quantum Information Center, University of California, Berkeley, California 94720}

\begin{abstract}
We study encodings that give the best known thresholds for the non-zero capacity of quantum channels, i.e., the upper bound for correctable noise, using an entropic approach to calculation of the threshold values.
%To explicitly obtain non-zero capacity codes, we further concatenate these encodings with an outer code; either a random code or with the $[[5,1,3]]$ code. 
%Using the [[5,1,3]] code, we show that after concatenating with this for a finite number of levels sufficient to obtain a small enough error, we can further construct an explicit encoding to arrive at a non-zero capacity code.    
Our results show that Pauli noise is correctable up to the hashing bound. For a depolarizing channel, this approach allows one to achieve a non-zero capacity for a fidelity (probability of no error) of $f=0.80870$.
\end{abstract}
\maketitle

This paper complements \cite{MoreChans}, which investigated how a given quantum error correcting code can best correct a particular type of noise. That work made use of an entropic approach to calculation of thresholds of correctable noise, showing that an adaptive concatenation of the quantum code can improve the thresholds.  In this paper, we apply the same entropic approach to finding the best known code for correcting a particular type of noise.  
%For a small enough value of the noise probability (e.g., 1\%), we can make an explicit construction for a non-zero capacity code, while for arbitrary noise probability we have to rely on the usual non-constructive use of concatenation with a random code in order to obtain the optimal finite capacity code. 
We find codes that can correct all Pauli noise up to the hashing bound, i.e., the error rate per bit when the Shannon entropy of the noise $S(\mathcal{N}) = 1$, disproving the conjecture \cite{SS} that there exists some uncorrectable Pauli noise below the hashing bound.

We first review the background to the current work.  From classical information theory, we know that the binary symmetric channel capacity (rate per bit at which information can be transmitted reliably using a large enough code) of some noise $\mathcal{N}$ is given as $Q_1 = 1 - S(\mathcal{N})$, where $S(\mathcal{N})$ is the Shannon entropy of the noise \cite{Shannon}. However, to get the true (optimal) quantum capacity of some noise, we must maximize the capacity over all codes of various lengths $n$, to arrive at a quantum capacity $Q = \lim_{n \rightarrow \infty} \frac{1}{n} Q_n \geq Q_1$ \cite{Lloyd,Devetak2003,ShorMSRI,SS,Schumacher}, where $Q_n$ is the maximum over a particular length. This paper is concerned with finding the threshold noise values at which the quantum channel capacity goes to zero: these thresholds constitute the limits of correctable noise. It had originally been conjectured that $Q = Q_1$ \cite{BDSW}, but it was subsequently found \cite{ShorSmolin,DiVincenzo98,SS} that some codes can correct above the noise values for which $Q_1=0$, implying that $Q > Q_1$.  We shall refer to the noise values for which $S(\mathcal{N})=1$ as the quantum Shannon hashing bound. Some codes that were shown to allow correction above the hashing bound consist of an $n_1$ qubit bit flip code concatenated with an $n_2$ qubit phase flip code \cite{DiVincenzo98,SS}.  We shall refer to these as ``$n_1$ in $n_2$'' codes, since each of the $n_2$ blocks of the phase flip code consists of an $n_1$ qubit bit flip code. In Ref. \cite{DiVincenzo98} these codes were concatenated with a random code to arrive at good codes that allow correction.  In this work we show that when these $n_1$ in $n_2$ codes are instead repeatedly concatenated with the $[[5,1,3]]$ code introduced in \cite{LMPZ}, we obtain the best known thresholds  for correcting quantum noise.  

\begin{figure}
\includegraphics[width=0.45\textwidth]{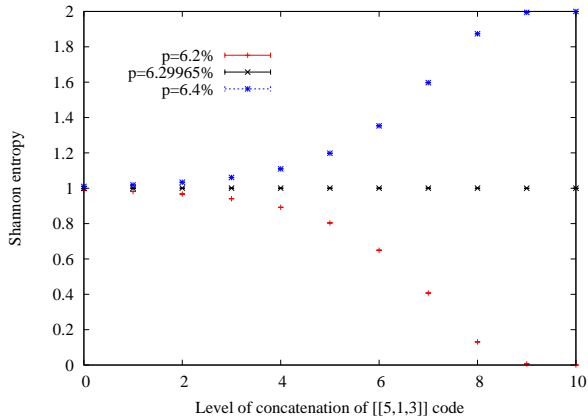}
\caption{\label{fig:513}
\capt Shannon entropy of depolarizing noise $p_X=p_Y=p_Z=p$ after variable levels of self-concatenation of the $[[5,1,3]]$ code, for noise at the threshold ($p \sim 6.29965 \%$) as well as somewhat below ($p = 6.2 \%$) and above ($p=6.4 \%$)  the threshold. These values are obtained from Monte Carlo simulation, with error bars considerably smaller than the plotted symbols.}
\end{figure}

The entropic approach to evaluation of the error threshold is as follows.  After we encode one logical qubit into $n$ physical qubits with a quantum error correcting code, we calculate the average of the Shannon entropy over all error syndromes, weighted by their probability of occurrence. If the average Shannon entropy of the logical noise is less than the hashing bound of $1$, then concatenation with a random code will result in a good code that allows complete correction of the noise \cite{DiVincenzo98}.
In this paper, we use repeated concatenation with a specific code (the $[[5,1,3]]$ code) rather than a random code.  The logical entropy is evaluated at successive levels of concatenation, using the procedure outline in \cite{MoreChans}.  The entropy as a function of concatenation level shows one of three behaviors, depending on the value of the error probability: i) for small error probabilities, $S(\mathcal{N})$ decreases with level $n$ to reach an asymptotic value of zero; ii) for one specific error probability value, following some initial variation with $n$, $S(\mathcal{N})$ goes to a constant value, usually approximately $1$; iii) for larger error probability values, $S(\mathcal{N})$ increases to an asymptotic value, which is always $2$ for the $[[5,1,3]]$ code, although it could be $1$ in special cases for other types of codes. 
Fig. \ref{fig:513} illustrates the average Shannon entropy as a function of concatenation level for depolarizing noise with the $[[5,1,3]]$ code concatenated repeatedly with itself.
The entropy versus concatenation level is shown for three different error probability values that illustrate the three generic behaviors: $p=0.062$ (below threshold), $p=0.629965$ (close to threshold) and $p=0.064$ (above threshold). 

In case i), the error can be corrected once the asymptotic value of zero is achieved, since the nature of the error is then known.  Case iii) corresponds to a combination of errors, with the maximum asymptotic value of 2 corresponding to depolarizing noise $(p_X, p_Y, p_Z) = (\frac{1}{4},\frac{1}{4},\frac{1}{4})$.  This channel results in the maximally mixed state, and cannot be corrected since there is no information on the nature of the error. Thus case ii), where the average Shannon entropy is asymptotically constant, provides the threshold error probability value of correctable noise, i.e., below the error probability for this value of $S(\mathcal{N})$, the noise can be corrected. An entropy of $0$ corresponds to a known correctable error, so the  level of concatenation at which zero entropy is first reached explicitly defines a code that can correct all errors up to the corresponding error probability threshold value. 
This procedure yields a code for a single error-free qubit.  
To achieve a non-zero capacity, we stop after a certain number of levels and then concatenate with a random code as in Refs.~\cite{DiVincenzo98,SS}.
% In Section~\ref{sec:constructive}, we discuss how these codes may then be used to obtain codes with non-zero quantum capacities.

The $n$ qubit bit flip code is the $n$ qubit repetition code which encodes $1$ logical qubit into $n$ physical qubits and which has encoded states $\overline{\ket{0}} = \ket{0}^{\otimes n}$ and $\overline{\ket{1}} = \ket{1}^{\otimes n}$. The $n$ qubit phase flip code has the encoded states $\overline{\ket{+}} = \ket{+}^{\otimes n}$ and $\overline{\ket{-}} = \ket{-}^{\otimes n}$, where $\ket{\pm} = \frac{1}{\sqrt{2}} \ket{0} \pm \frac{1}{\sqrt{2}} \ket{1}$. We now consider the $n_1$ in $n_2$ code. If this were concatenated repeatedly with itself, the resulting logical noise would tend to be unbalanced in terms of phase errors and bit flip errors, and therefore tend to be dominated by either $X$ or $Z$ errors.
In this situation we have found that it is hard to calculate the thresholds for which the entropy is unity using Monte Carlo, since this point becomes an attracting fixed point for the bit flip and phase flip codes.

Instead, we use the $n_1$ in $n_2$ code as input to repeated concatenation with the $[[5,1,3]]$ code.
Our rationale for this procedure is the fact that the $[[5,1,3]]$ code is also very efficient at correcting noise dominated by one type of error. In particular, if there is probability $p$ of a single Pauli error 
$\sigma$ and probability $1-p$ of no error, with the exception of the special case that when $p=50\%$,  the $[[5,1,3]]$ code can always correct the error. To illustrate this, Tab. \ref{tab:dom} shows an example of the performance of a self-concatenated $[[5,1,3]]$ code for $X$-dominated error, compared to that of a self-concatenated $5$ in $5$ code.  We see that both codes can correct noise dominated by $X$ errors above the hashing bound and that the thresholds for the $[[5,1,3]]$ code are just below that of the combined bit flip/phase flip 5 in 5 code.  Similar results are found for noise dominated by $Z$ errors.

\begin{table}
\caption{\label{tab:dom} Upper bound on correctable noise, i.e., threshold value, for Pauli noise channel $(p_X,p_Y,p_Z) = (p,10^{-6},10^{-6})$. By definition, the hashing bound is the threshold value for level 0, i.e., no concatenation.} 
\begin{ruledtabular}
\begin{tabular}{cccc}
Level & [[5,1,3]] & $5$ in $5$ code \\
\hline
0 & 49.62410483\% & 49.62410483\% \cr
1 & 49.64614794\% & 49.64614908\% \cr
2 & 49.66961046\% & 49.66961385\% \cr
\end{tabular}
\end{ruledtabular}
\end{table}
  
An important consequence of repeatedly concatenating the initial bit flip/phase flip $n_1$ in $n_2$ code with the $[[5,1,3]]$ code, is that when the noise is below the threshold value, the logical entropy converges to $0$ with increasing concatenation level. This means that the logical noise converges to a single Pauli error and can be corrected, as outlined above and demonstrated in \cite{MoreChans}. Therefore this combination of $n_1$ in $n_2$ followed by concatenation with $[[5,1,3]]$ gives a constructive approach for correcting all noise below the threshold.  
To compute the thresholds from repeated concatenation, we first estimate the threshold after the $n_1$ qubit bit flip code is followed by the $n_2$ qubit phase flip code and then by some fixed number of levels of the $[[5,1,3]]$ code.  This fixed number is chosen to maximize the lower part of the confidence interval in the Monte Carlo evaluation of thresholds (see below for a discussion of the Monte Carlo simulations). The true threshold value is then obtained from further repeated concatenation with the $[[5,1,3]]$ code, resulting in a slightly higher value than the value $p \sim 0.062995$ shown in Fig.~\ref{fig:513}.

If the probabilities of Pauli errors $\sigma$ occurring 
on a single qubit are $p_{\sigma}$, the Shannon entropy is 
\begin{equation*}
S(\mathcal{N}) = H(p_I, p_X, p_Y, p_Z) = \sum_{\sigma} h(p_{\sigma}),
\end{equation*}
where 
\begin{equation*}
h(x) = -x \log_2 x.
\end{equation*} 
$p_X$ is the probability of a pure bit flip error (without a phase error), $p_Z$ the probability of a pure phase error (without a bit flip error) and $p_Y$ the probability of a combined bit flip and phase flip error. The fidelity $f$ is defined to be the probability of no error, i.e., $p_I$. 
We now consider how to construct the entropy for logical errors in encoded qubits, first without concatenation and then with repeated concatenation.

\begin{lemma}
Suppose that there is identical independent noise on each qubit of an $n$ qubit bit flip code, with $p_\sigma$ representing the probability of a $\sigma$ Pauli error. The probability of measuring a syndrome that corresponds to an error of distance $k$ and associated logical $\sigma$ error is $l_{\sigma}(k)$, where
\begin{align}
\label{eq:LogicalErrors}
l_I(k) = \frac{a_k + b_k}{2} &&
l_X(k) = \frac{a_{n-k}+b_{n-k}}{2} \nonumber \\
l_Y(k) = \frac{a_{n-k}-b_{n-k}}{2} &&
l_Z(k) = \frac{a_k - b_k}{2} 
\end{align}
and 
\begin{align*}
a_k = \binom{n}{k} q_X^k(1-q_X)^{n-k} \\
b_k = \binom{n}{k} (p_X-p_Y)^k (p_I-p_Z)^{n-k}.
\end{align*}
For the special case of $k=\frac{n}{2}$, due to double counting we have
\begin{align}
\label{eq:specialcase}
l_I(\frac{n}{2}) = l_X(\frac{n}{2}) = \frac{a_{\frac{n}{2}}+b_{\frac{n}{2}}}{4} \nonumber\\
l_Y(\frac{n}{2}) = l_Z(\frac{n}{2}) = \frac{a_{\frac{n}{2}}-b_{\frac{n}{2}}}{4}.
\end{align}
\end{lemma}
Here $q_X$ represents the total probability of some sort of bit flip error (that is an $X$ or a $Y$) Pauli error. Therefore, $q_X = p_X+p_Y$. 
\begin{proof}
The probability of no logical bit flip error, $l_I(k) + l_Z(k)$, is the binomial distribution of $q_X$, which is defined above as $a_k$. The complement of this is the probability of a logical bit flip error, $l_X(k) + l_Y(k) = a_{n-k}$.

Suppose we have two independent noise sources for phase errors on single qubits, with probabilities $p_1$ and $p_2$ of occurring. Then the total resulting probability of an error is $p = \frac{1 - (1-2p_1)(1-2p_2)}{2}$, which is a classical result.  

In the following, we derive all four logical error probabilities $l_\sigma(k)$ by analyzing the probabilities of logical phase errors. First we look at the case where there is a logical phase error but no logical bit flip error. On $k$ qubits, there are no bit flip errors, and $p_1 = \frac{p_Z}{p_X+p_Z}$ is the conditional probability of a phase error. On the other $n-k$ qubits, there are bit flip errors, and $p_2 = \frac{p_Y}{p_X+p_Y}$ is the conditional probability of a phase error. Thus the conditional probability of a logical phase error is
\begin{align*}
\frac{l_Z(k)}{l_I(k)+l_Z(k)} = \frac{1 - (1 - 2p_1)^k (1-2p_2)^{n-k}}{2}
\\= \frac{1}{2} - \frac{1}{2} (\frac{p_X - p_Y}{p_X + p_Y})^k (\frac{p_I-p_Z}{p_I + p_Z})^{n-k} = \frac{1}{2} - \frac{1}{2} \frac{b_k}{a_k}.
\end{align*}
Using the fact that $l_I(k) + l_Z(k)=a_k$, we can eliminate $l_I(k)$ to obtain $l_Z(k) = \frac{a_k - b_k}{2}$ and hence $l_I(k) = \frac{a_k + b_k}{2}$.

Now we examine the case when there are both a logical bit flip error and a logical phase flip error.  When we have distance $k$ logical phase flip and bit flip errors together, the conditional logical probability of a distance $k$ phase flip error is $\frac{l_Y(k)}{l_X(k)+l_Y(k)} $.
We note that to cause a logical bit flip error of distance $k$ ($n-k$), there must actually be $n-k$ ($k$) bit flip errors. Therefore, this conditional probability can also be written as the logical probability of a distance $n-k$ phase flip given no $n-k$ bit flip errors, $\frac{l_Z(n-k)}{l_I(n-k)+l_Z(n-k)}$.  Hence we have
\begin{equation*}
\frac{l_Y(k)}{l_X(k)+l_Y(k)} = \frac{l_Z(n-k)}{l_I(n-k)+l_Z(n-k)}.
\end{equation*}
Using the fact that $l_X(k)+l_Y(k)=a_{n-k}$, this gives $l_Y(k)=\frac{a_{n-k}-b_{n-k}}{2}$ and therefore $l_X(k) = \frac{a_{n-k}+b_{n-k}}{2}$. 
\end{proof}

\begin{corollary}
\label{cor:averageEntropy}
The total logical entropy resulting from an $n$ qubit bit flip code is
\begin{equation*}
%\label{eq:averageEntropy}
\sum_{k=0}^{\frac{n}{2}}  \big( l(k) \sum_{\sigma} h(\frac{l_{\sigma}(k)}{l(k)}) \big) = \sum_{k=0}^{\frac{n}{2}} \big( - h(l(k)) + \sum_{\sigma} h(l_{\sigma}(k)) \big) ,
\end{equation*}
where $l(k) = \sum_\sigma l_\sigma(k)$ is the probability of a syndrome corresponding to a distance $k$ error. $l(k) = a_k + a_{n-k}$ except in the special case of $k=\frac{n}{2}$, where $l(\frac{n}{2}) = a_{\frac{n}{2}}$
because of double counting (see also Eq.~\ref{eq:specialcase}).
\end{corollary}
\begin{proof}
The conditional probability of having a logical $\sigma$ error, given that there was a distance $k$ error measured, is $\frac{l_\sigma(k)}{l(k)}$. Weighted by the probability $l(k)$ of a distance $k$ error being measured, this contributes $l(k) h(\frac{l_\sigma(k)}{l(k)})$ to the total entropy. Summing over all $k$ and $\sigma$ gives the desired result.
\end{proof}

To show how the logical entropy is then evaluated after the code is concatenated we use the example of an $n_1=5$ bit flip code concatenated with an $n_2$ qubit phase flip code.  First, we calculate the values of $l_\sigma(k)$ for the $3$ error syndromes corresponding to distance $k=0,1,2$ errors. There are $3$ corresponding types of sub-blocks, each with a distance $0$, $1$, or $2$ error. For each of these $3$ types of sub-blocks, the $n_2$ qubit phase flip code can either have an error detected on that block or not, resulting in $v=6$ different possible error cases for a block. Second, a multinomial expansion of these $6$ cases is performed. This allows for exact calculations for the $n_1=5$ in $n_2$ code thresholds. In general, the number of cases to consider is the number of ways for $v$ non-negative integers to sum to $n_2$, which is approximately $\frac{n_2^{v-1}}{(v-1)!}$. For odd $n_1$, there are $v=n_1+1$ cases. The most time intensive exact calculation performed for this paper was for the $7$ in $134$ code: this requires approximately $1.5 \times 10^{11}$ steps, a calculation that takes on the order of weeks on typical 2008 CPU.  The Monte Carlo simulations on larger systems described below took up to $5 \times 10^{12}$ steps, with the largest simulation taking on the order of months on a typical 2008 CPU.

Some calculations in this paper use a Monte Carlo method to estimate the error thresholds, in particular, the calculations for the doubly concatenated codes having an additional $j$ levels of concatenation with the $[[5,1,3]]$ code.  The threshold values are defined as the noise values for which the entropy is a constant value (usually $1$, as discussed above) for a given $n_1$ qubit bit flip code in a $n_2$ qubit phase flip code, which is then further concatenated with the $[[5,1,3]]$ code. At the first level of encoding (it need not be the same code at each level), the same noise is applied on each qubit. At each block of each level, a syndrome is chosen randomly from the error syndromes, weighted by their probability of occurrence. The resulting probabilities of logical errors $l_\sigma$ are computed, and the logical noise from each block of one level is passed on as noise on a single qubit to the next level of the code. At the top level of the code, the total logical errors $l_\sigma$ are computed, and the resulting Shannon entropy of the errors $H(l_I, l_X,l_Y,l_Z)$ then computed. This entropy is then averaged over many Monte Carlo samples of the first level logical error probabilities to obtain a statistical estimate of the Shannon entropy of the total logical noise. (The same Monte Carlo method was used in \cite{MoreChans}.) The simulations are run for highly divisible values of $n_2$, by treating the $n_2$ phase flip code as several phase flip codes concatenated with themselves, since there is a lot of overhead associated with having a large code at one level of the encoding.  All results obtained with Monte Carlo are shown with the standard error, which corresponds to a $68\%$ confidence interval. 

For an example of how the concatenation of syndromes works, suppose that we have the $n_1=2$ qubit bit flip code concatenated into the $n_2=3$ qubit phase flip code. Now suppose that the error recovery operator for the first block is $I \otimes I$, and for the second and third blocks is $I \otimes X$ in each case. Then after the first code (the $2$ qubit bit flip code), the total recovery operator on the $6$ qubits is $I \otimes I \otimes I \otimes X \otimes I \otimes X$. Now, suppose that the recovery operator for the phase flip code is $I \otimes I \otimes Z$, and encoded $\overline{Z}$ on the bit flip code is $\overline{Z}=I \otimes Z$. This multiplies a recovery operator of $I^{\otimes 5} \otimes Z$, which resulting in a final combined recovery operator of $I \otimes I \otimes I \otimes X \otimes I \otimes Y$. 

\begin{table}
\caption{\label{tab:5-64} 
Threshold values for noise correction by the $5$ qubit bit flip code concatenated $6$ times with the $2$ qubit phase flip code (resulting in the $5$ in $2^6$ code after $7$ levels), and then concatenated repeatedly with the $[[5,1,3]]$ code. We find the values of $p$ for which the logical entropy is $1$ at each level for different types of errors $(p_X,p_Y,p_Z)$.  $(p,p,p)$ is the depolarizing channel, $(p-p^2,p^2,p-p^2)$ represents bit flip and phase flip errors occurring independently, and $(p,0,p)$ is two-Pauli noise.  
For the lower levels, the threshold calculations may be performed exactly, by numerical inversion of the exactly calculated average Shannon entropy.  For higher levels, the rapidly increasing number of syndromes renders exact calculation unacceptably inefficient and the entropy is then evaluated by Monte Carlo sampling (see text).  For these calculations the standard error corresponding to a 68\% confidence interval is shown in standard form as the last digit in parentheses.
}
\begin{ruledtabular}
\begin{tabular}{llll}
Level & $(p,p,p)$   & $(p-p^2,p^2,p-p^2)$    & $(p,0,p)$\\
\hline
0           & 6.30965616\% & 11.00278644\% & 11.35460976\%\cr
1           & 6.34520294\% & 11.21042175\% & 11.33392680\%  \cr
2           & 6.34750308\% & 11.21331544\% & 11.33595709\% \cr
3           & 6.35074316\% & 11.21812585\% & 11.33994152\% \cr
4           & 6.35541320\% & 11.22592213\% & 11.34718019\% \cr
5           & 6.36255660\% & 11.23929192\% & 11.36035645\% \cr
6           & 6.37084591\% & 11.25886132\% & 11.37968385\% \cr
7           & 6.37272029\% & 11.27375652\% & 11.39372640\% \cr
8           & 6.373(5)\%   & 11.275\%      & 11.395(1)\% \cr
9           & 6.375\%      & 11.277\%      & 11.397\% \cr
10          & 6.376\%      & 11.27(7)\%    & 11.398\% \cr
$\infty$    & 6.376\%      & 11.27(8)\%    & 11.39(8)\% \cr
\end{tabular}
\end{ruledtabular}
\end{table}

In Tab. \ref{tab:5-64}, we show noise threshold values obtained for the $5$ qubit bit flip code concatenated $6$ times with the $2$ qubit phase flip code to make a $5$ in $64$ code which is further concatenated repeatedly with the $[[5,1,3]]$ code. Calculations were made for three types of Pauli noise: depolarizing, independent bit and phase flips, and two-Pauli noise. At each level, we determine the error probabilities $p_\sigma$ that give a logical entropy of $1$. 
For the three types of noise considered, the thresholds found here are close to the optimal values found later in this paper.

\paragraph{Example}
Suppose that we wish to know the entropy contribution from the syndrome of distance $k=1$ for the $n=2$ qubit bit flip code, under depolarizing noise $p_X=p_Y=p_Z=p$. In Eq. \ref{eq:LogicalErrors}, $a_1 = 4p - 8p^2$, $b_1=0$. Since $k=\frac{n}{2}$, we have the special case where dividing by $2$ from Eq. \ref{eq:LogicalErrors} (see discussion below Eq. \ref{eq:LogicalErrors}) gives us $l_\sigma(1) = p - 2p^2$ for all $\sigma$. From Corollary \ref{cor:averageEntropy}, the entropy for the $k=1$ syndrome is therefore 
\begin{equation*}
l(1) \sum_\sigma h(\frac{l_\sigma(1)}{l(1)}) = 4 l(1) h(\frac{1}{4}) = 4 l(1) \frac{1}{2} = 8p-16p^2.
\end{equation*}

The codes presented above are examples of doubly concatenated codes, where the first concatenation constructing the $n_1$ in $n_2$ code is finite and the second, subsequent concatenation of this with the $[[5,1,3]]$ code is unlimited.  The latter concatenation allows the average Shannon entropy of the logical errors and hence the rate of unknown errors to be driven to zero, while the first, finite concatenation provides an additional parameter, $n_2$, that can be optimized to find the highest possible threshold values.

In Sections \ref{sec:depol} - \ref{sec:2Pauli} we present examples of optimization of these doubly concatenated codes for several different Pauli noise models described by one or two noise parameters.  In Section \ref{sec:infinite} we present threshold results for correction of independent noise in all three Pauli degrees of freedom, using an infinite length bit flip code. %In Section \ref{sec:constructive}, we discuss our claim of a constructive method yielding a non-zero capacity. 

\begin{figure*}
\includegraphics[width=1.00\textwidth]{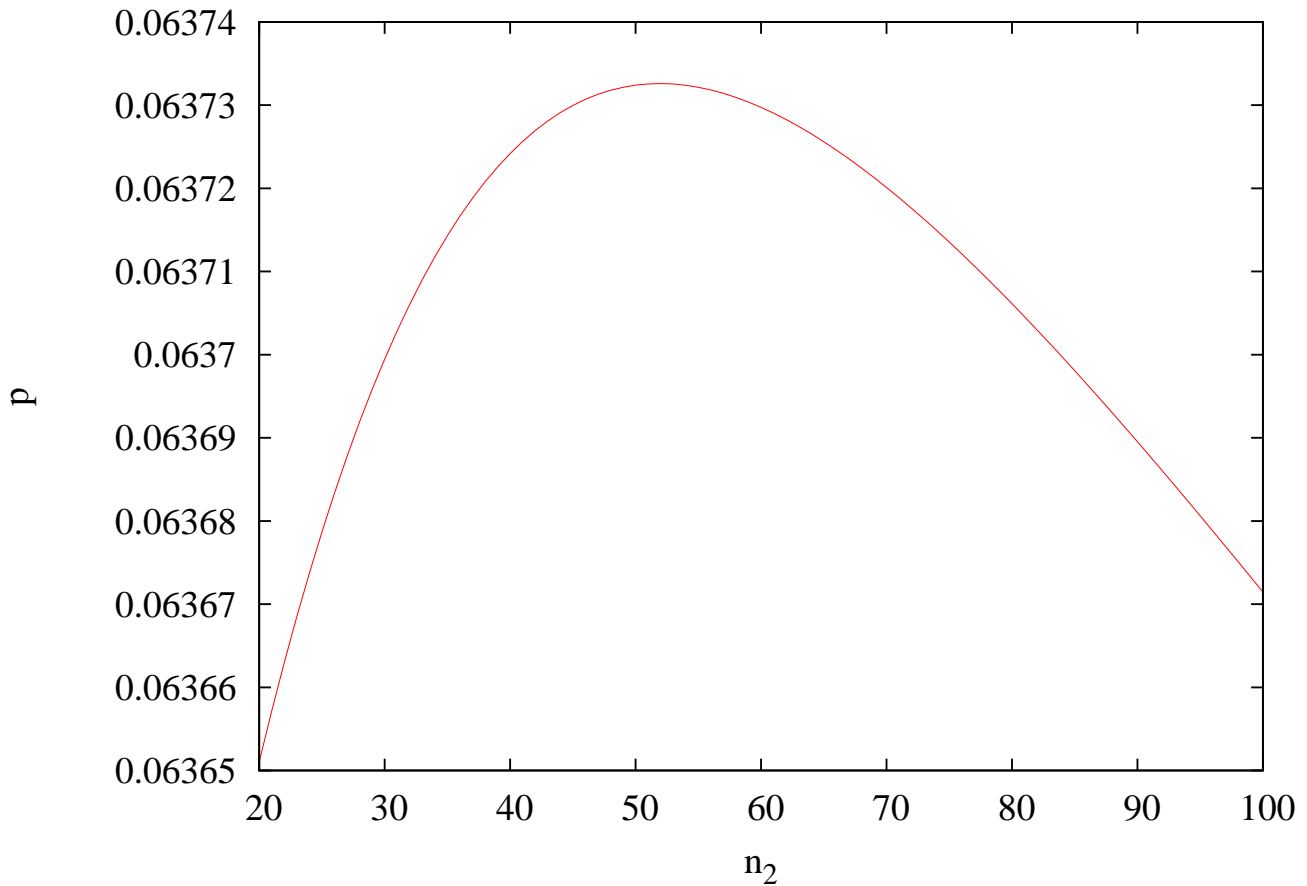}
\caption{\label{fig:depolarvalues}
\capt Thresholds of non-zero capacity for the depolarizing channel for $5$ qubit bit flip code concatenated with $n_2$ qubit phase flip code. These values are calculated exactly.}
\end{figure*}

\begin{table}
\caption{\label{tab:bitflip} 
Values of $p$ for which the logical entropy is $1$ for an $n_1$ qubit bit flip code for  noise $(p_X,p_Y,p_Z)$. Note that here the values for $n_1=1$ correspond to the hashing bound. These values are all calculated exactly.
}
\begin{ruledtabular}
\begin{tabular}{cccc}
$n_1$          & $(p,p,p)$  & $(p-p^2,p^2,p-p^2)$ & (p,0,p) \cr
\hline          
1          & 6.30965616\%   &  11.00278644\%  &  11.35460976\%\cr 
2          & 6.28410724\%   &  11.00278644\%  &  11.18454296\%\cr
3          & 6.33766430\%   &  11.16520399\%  &  11.30915446\%\cr
4          & 6.32983488\%   &  11.16162540\%  &  11.29120242\%\cr
5          & 6.34520293\%   &  11.21042175\%  &  11.33392680\%\cr
6          & 6.33623898\%   &  11.19383617\%  &  11.31378370\%\cr
7          & 6.34108373\%   &  11.21074102\%  &  11.32891165\%\cr
8          & 6.33195564\%   &  11.19067373\%  &  11.30752673\%\cr
9          & 6.33268543\%   &  11.19549408\%  &  11.31166177\%\cr
$\infty$   & 6.06394190\%   &  10.69243112\%  &  10.79171085\%\cr
\end{tabular}
\end{ruledtabular}
\end{table}

\section{Depolarizing noise}

\begin{table}
\caption{\label{tab:depo} 
For a given $n_1$ qubit bit flip code, we find the optimal $n_2$ phase flip code for correcting depolarizing noise, i.e. $(p_X,p_Y,p_Z)=(p,p,p)$. The threshold values are determined to be the value of $p$ for which the entropy is equal to $1$. The hashing bound (corresponding to the $1$ in $1$ code) is $p=6.30965616\%$.  These results are determined exactly.}
\begin{ruledtabular}
\begin{tabular}{ccc}
$n_1$      & Optimal $n_2$   & Threshold \cr
\hline
1          &  5   & 6.34520293\% \cr
2          &  1   & 6.28410724\% \cr
3          &  19  & 6.36189692\% \cr
4          &  1   & 6.32983488\% \cr
5          &  51  & 6.37338273\% \cr
6          &  11  & 6.34136778\% \cr
7          &  133 & 6.36907054\% \cr
8          &  38  & 6.34112748\% \cr
\end{tabular}
\end{ruledtabular}
\end{table}

\label{sec:depol}
Depolarizing noise is noise of the form $(p_X,p_Y,p_Z)=(p,p,p)$. For a bit flip code with $n_1$ qubits, values of $p$ for which the logical entropy is $1$ are given in Tab. \ref{tab:bitflip}. Previous work \cite{DiVincenzo98} has shown that $n_1=5$ is optimal for this class of codes (termed 'cat' codes in \cite{DiVincenzo98} and 'repetition' codes in \cite{SS}), and that this code also corrects above the hashing bound. 

Now we consider an $n_1$ qubit bit flip code concatenated with an $n_2$ qubit phase flip code. Previous work \cite{SS} has shown that for $n_1=3$, the optimal length for the phase flip code is $n_2=19$, and yields a vanishing logical error rate at the threshold value $p=6.32\%$, corresponding to a logical entropy of 1.
Smith and Smolin \cite{SS} were also able to evaluate the threshold for $n_1=5$ up to $n_2 = 16$ but not beyond this.  However, for the $n_1$ in $n_2$ codes it is possible to make efficient, exact calculations of the average entropy and threshold values.  Fig. \ref{fig:depolarvalues} shows the threshold values resulting from such exact calculations as a function of $n_2$ for the $n_1=5$ qubit bit flip code concatenated with an $n_2$ qubit phase flip code.  It is evident from this plot that the value $n_2 = 51$ is optimal, with a corresponding threshold value of $p=6.37338273\%$, with a corresponding channel fidelity of  $f= 1-3p = 0.80879852$.  
Tab. \ref{tab:depo} shows the maximal thresholds as a function of $n_2$ for values of $n_1 \leq 8$.  These results show that the highest threshold is obtained for $n_1=5, n_2=51$, by a considerable margin, and suggest that this code is therefore optimal for all $n_1, n_2$.

As expected from the discussion in the previous section, it turns out that we can do even better by concatenating $5$ in $n_2$ codes repeatedly with the $[[5,1,3]]$ code. For this doubly concatenated code we need to make use of Monte Carlo sampling to evaluate the average Shannon entropy at the higher concatenation levels of the $[[5,1,3]]$ code.  We find that upon concatenating the $5$ in $56$ code $10$ times with the $[[5,1,3]]$ code, the threshold value is $p=6.376753\% \pm 0.000006\%$, corresponding to a fidelity value of $0.80869740 \pm 0.00000018$.   
Tab. \ref{tab:compare} compares this threshold value for the depolarizing channel with the various $5$ in $n_2$ codes discussed above and shows that the doubly concatenated code provides an improvement over all other existing codes.

\begin{table}
\caption{\label{tab:compare} 
Comparison of thresholds found for the depolarizing channel $p_X=p_Y=p_Z=p$ with the doubly concatenated code ($5$ in $56$ repeated $[[5,1,3]]$) with the optimal $5$ in $n_2$ code and with previous results for other $5$ in $n_2$ codes \cite{DiVincenzo98, SS} and for the repetition code ($5$ qubit bit flip \cite{ShorSmolin}). The threshold values for the doubly concatenated code are evaluated with the use of Monte Carlo sampling, and the statistical error is given as the standard error corresponding to a 68\% confidence interval. The other threshold values are evaluated exactly.}
\begin{ruledtabular}
\begin{tabular}{cll}
Code         & $p$            & fidelity         \cr
\hline 
Hashing rate                     & 6.30965616\%  & 0.81071032 \cr
$5$ qubit bit flip \cite{ShorSmolin}   & 6.34520293\%  & 0.80964391 \cr
$5$ in $5$ \cite{DiVincenzo98}   & 6.35204743\%  & 0.80943858 \cr
$5$ in $16$ \cite{SS}            & 6.36255660\%  & 0.80912330 \cr
$5$ in $51$                      & 6.37338273\%  & 0.80879852 \cr
$5$ in $56$ repeated $[[5,1,3]]$ & 6.3767(5)\%   & 0.80869(7) \cr
\end{tabular}
\end{ruledtabular}
\end{table}

\begin{table}
\caption{\label{tab:ind} 
For a given $n_1$ qubit bit flip code, we find the optimal $n_2$ phase flip code for correcting noise of the form $(p_X,p_Y,p_Z)=(p-p^2,p^2,p-p^2)$, i.e., independent noise at the unique set of parameter values $q_X = q_Z = p$. The threshold values are determined to be the value of $p$ for which the entropy is equal to $1$. The hashing bound (corresponding to the $1$ in $1$ code) is $p=11.00278644\%$.  These results are determined exactly.}
\begin{ruledtabular}
\begin{tabular}{ccc}
$n_1$      & Optimal $n_2$   & Threshold \cr
\hline
1          &  7   & 11.21074102\% \cr
2	        &  1   & 11.00278644\% \cr
3          &  25  & 11.23097281\% \cr
4          &  1   & 11.16162540\% \cr
5          &  77  & 11.27458434\% \cr
6          &  12  & 11.20118393\% \cr
7          &  221 & 11.27420360\% \cr
8          &  52  & 11.20675588\% \cr
\end{tabular}
\end{ruledtabular}
\end{table}

\section{Independent noise}
\label{sec:indep}
Suppose that we have independent rates $q_X$ of an $X$ or $Y$ Pauli error (some sort of bit flip), and $q_Z$ of a $Y$ or $Z$ Pauli error (some sort of phase flip). Written in terms of Pauli error probabilities, this is $(p_X,p_Y,p_Z)=(q_X(1-q_Z),q_X q_Z, q_Z(1-q_X)$. Because they are detected independently of each other, the bit flips and phase flips can be corrected independently of each other. Therefore, information about the bit flip errors doesn't affect the correction of the phase flip errors, and vice versa.

We apply an $n_1$ qubit bit flip code concatenated with an $n_2$ qubit phase flip code. Let the binary entropy function be $f_b(p) = h(p) + h(1-p)$. For the $X$ entropy, if we detect a distance $k$ error, we have either a bit flip error with probability $a_{n_1-k}$, or no bit flip error with probability $a_k$. We perform an $n_2$ degree multinomial expansion to get the total $X$ entropy. After the bit flip code, the rate of phase flips is $q_Z' = \frac{1-(1-2q_Z)^{n_1}}{2}$. If $a_k = \binom{n_2}{k} q_Z'^k (1-q_Z')^{n_2-k}$, the $Z$ entropy is $Z_e = \sum_k (a_k + a_{n_2-k}) f_b(\frac{a_k}{a_k+a_{n_2 -k}})$. 

Making use of the results of Prop. \ref{prop:infinitenoise}, we have
%\begin{equation*}
$2 (q_X(1-q_X))^{\frac{3}{2}} < (q_X (1-2 q_Z))^2 (1-q_X) + ((1-q_X)(1-2 q_Z))^2 q_X$,
%\end{equation*}
which gives a threshold of
\begin{equation*}
q_X = \frac{1 - \sqrt{1 - (1 - 2 q_Z)^4}}{2}.
\end{equation*}
For $q_X=q_Z$, the critical value is $q_X=q_Z = \frac{1}{2} - \frac{\sqrt{2 \sqrt{5}-2}}{4}$. For small $q_Z$, the optimal $n_1$ is approximately $n_1 \approx \frac{1}{2q_Z}$, and the threshold is $(p_X,p_Y,p_Z)=(\frac{1}{2}-\epsilon,\frac{1}{4}\epsilon^2,\frac{1}{2}\epsilon^2)$.

Tab. \ref{tab:ind} shows the $n_2$-optimized threshold values for $n_1$ ranging from 1 to 8, in the case that the independent noise parameters are all equal, i.e.,  $q_X = q_Z = p$.   We find that the $5$ in $77$ code is best at correcting this type of noise, with a threshold value of $11.27458434\%$.  This result is somewhat surprising, since this special case of independent noise is the same as that in the second column of Tab. \ref{tab:bitflip} for pure bit flip codes, i.e., $n_2=1$.  In the latter case we found that the $7$ qubit bit flip code performed slightly better than the $5$ qubit bit flip code.  In contrast, Tab. \ref{tab:ind} shows that when the bit flip code is concatenated with a phase flip code so that $n_2 > 1$, the $5$ qubit bit flip code performs slightly better than the $7$ qubit bit flip code.
However, we find that we can do even better by repeatedly concatenating with the $[[5,1,3]]$ code. For example, for the $5$ in $84$ code concatenated $9$ times with the $[[5,1,3]]$ code, a threshold of $11.28061\% \pm 0.00005\%$ is calculated, which is better than the optimal $n_1$ in $n_2$ code of Tab. \ref{tab:ind}.

\begin{figure*}
  \begin{minipage}[b]{0.48\textwidth}
    \includegraphics[width=\textwidth]{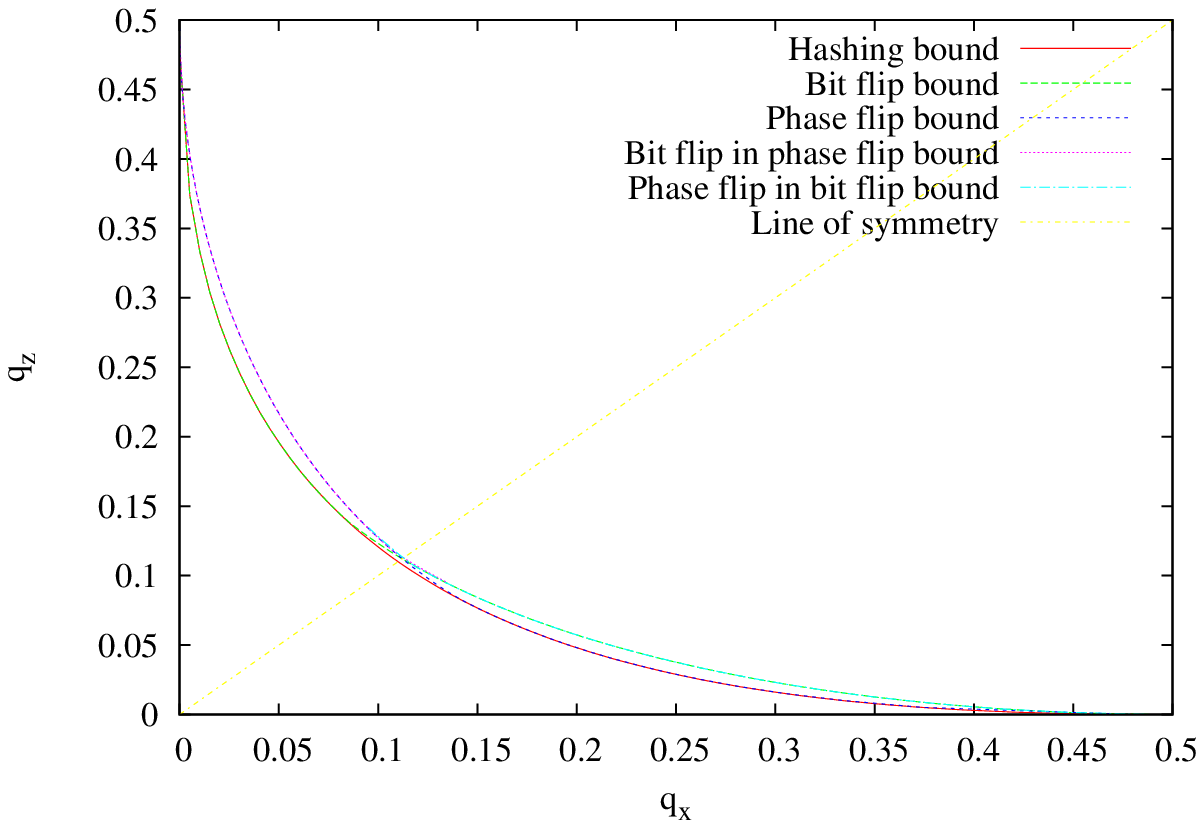}
  \end{minipage}
  \begin{minipage}[b]{0.48\textwidth}
    \includegraphics[width=\textwidth] {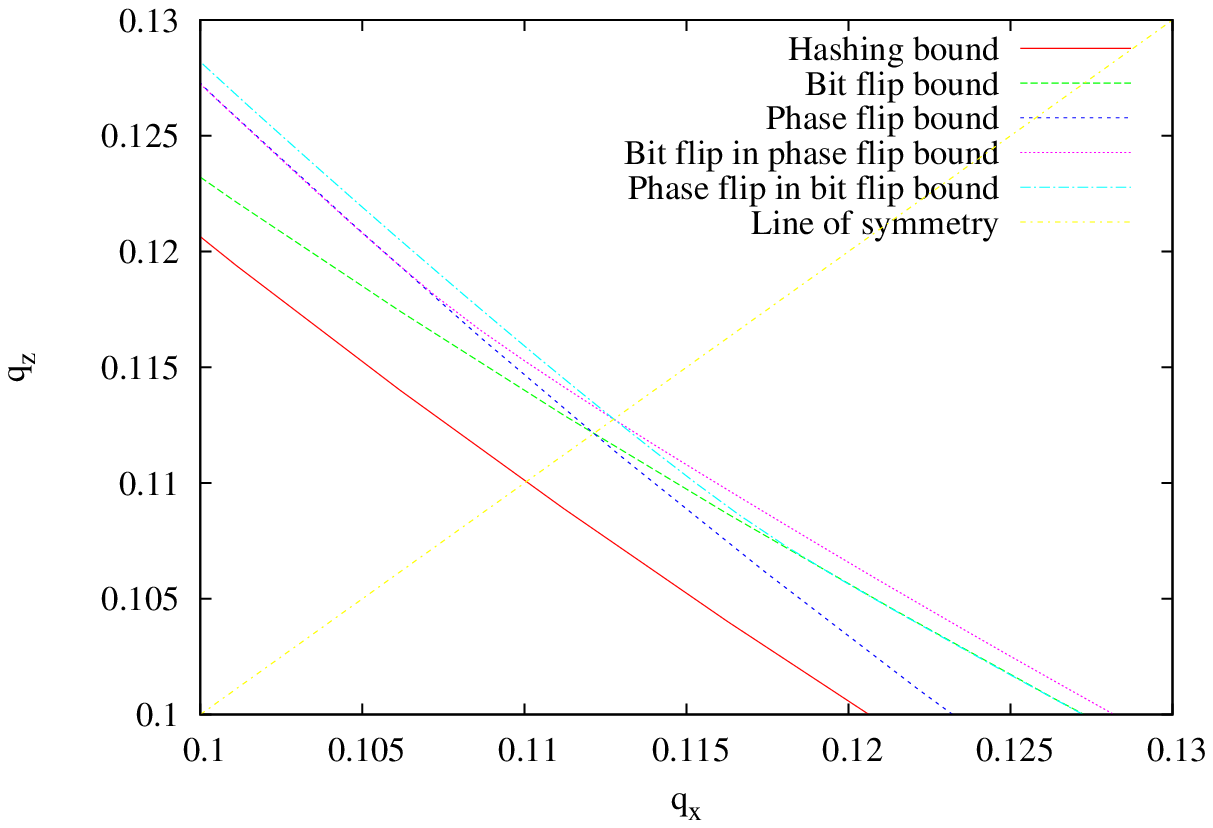}  
  \end{minipage}

\caption{\label{fig:ind} 
\capt Threshold of the non-zero capacity for independent noise $(p_X,p_Y,p_Z)=(q_X(1-q_Z),q_X q_Z, q_Z(1-q_X))$. The yellow line shows the line of symmetry where $q_X = q_Z$ and the red line the hashing bound for which the noise entropy is 1. The bit flip (green line) and phase flip (blue line) bounds use the optimal bit flip and phase flip code, respectively (optimization was carried out up to $n \leq 1500$).The bit flip in phase flip (purple line) and phase flip in bit flip (turquoise line) bounds use the optimal $n_1$ in $n_2$ code for $n_1 < 9$ and $n_2 \leq 219$. The bit flip and phase flip bounds cross at $q_Z=q_Z=11.21074102\%$, which is the threshold for the $n_1=7$ code when $q_X=q_Z$, and the bit flip in phase flip and phase flip in bit flip bounds cross at $q_X=q_Z=11.27458434\%$, which is the threshold for the $5$ in $77$ code when $q_X=q_Z$ (see Tab. \ref{tab:ind}).  The right hand panel shows an enlargement of the crossover region around the symmetry line.
}
\end{figure*}

 Fig. \ref{fig:ind} shows a plot of the two-dimensional thresholds ($q_X$, $q_Z$) for the general independent noise situation, with four different types of $n_1$ in $n_2$ encodings. The hashing rate at which the noise entropy is 1 is shown as a red line for reference. The line of symmetry (brown line) is $q_X=q_Z$. We divide the threshold behavior into two regions, before the symmetry line, $q_X < q_Z$, and beyond the symmetry line, $q_X > q_Z$. Before the symmetry line, i.e., for $q_X < q_Z$, we find that among these four codes, it is optimal to apply a phase flip code first and then a bit flip code (which is reversed from an $n_1$ in $n_2$ code). After the symmetry line, i.e., for $q_X > q_Z$, the optimal of these types of codes is to apply a bit flip code first and then a phase flip code (i.e., the usual $n_1$ in $n_2$ code).  Fig. \ref{fig:ind} shows that the bit flip and phase flip bounds meet and cross at the symmetry line, as do the bit flip in phase flip and phase flip in bit flip bounds.

\section{Two-Pauli noise}
\label{sec:2Pauli}
We now consider noise of the form $(p_X,p_Y,p_Z)=(p_1,p_2,p_1)$. We refer to this as two-Pauli noise. It is known that bit flip codes by themselves cannot correct two-Pauli noise up to the hashing bound near $(p,0,p)$ \cite{SS}. Tab. \ref{tab:bitflip} demonstrates that the $5$ qubit bit flip code is the optimal bit flip code for correcting in this region. 

\begin{figure}
%\begin{center}
\includegraphics[width=0.45\textwidth]{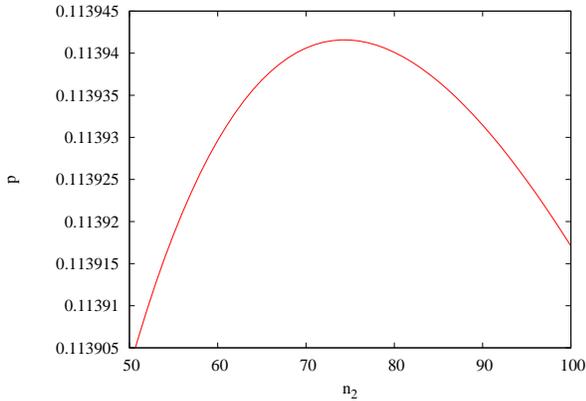}
%\includegraphics{two}
%\end{center}
\caption{\label{fig:two}
\capt Threshold of the non-zero capacity for $(p_X,p_Y,p_Z)=(p,0,p)$ channel for $5$ qubit bit flip code concatenated with a $n_2$ qubit phase flip code. These values are determined exactly.}
\end{figure}

As can be seen in Fig. \ref{fig:two}, we can correct above the hashing bound of $p=11.35460976\%$ for $(p,0,p)$ type noise by concatenating the $n_1=5$ qubit bit flip code with an $n_2$ qubit phase flip code. The $5$ in $74$ code is optimal and yields a threshold of $p=11.39425214\%$. The $5$ in $n_2 \approx 84$ code concatenated $8$ times with the $[[5,1,3]]$ code yields a threshold of $11.40030\% \pm 0.00006\%$.

Previous work \cite{SS} found that there always exists some bit flip or phase flip code that can correct up to at least the hashing bound for all Pauli noise, except for a neighborhood of this $(p,0,p)$ type noise. It had been conjectured \cite{SS} that a neighborhood of this point was not correctable. We have performed exact calculations which show that for physical noise at the hashing bound in this region such that $p_X \geq p_Z$,  the $5$ in $16$ code (i.e., the $5$ qubit bit flip code in $16$ code phase flip code) has a logical entropy of at most $0.9990331$ (and goes to $0$ under repeated concatenation with the $[[5,1,3]]$ code); therefore it can correct above the hashing bound in this segment of the region.  The same applies to the symmetrical $5$ qubit phase flip code in $16$ qubit bit flip code for the complementary segment $p_X \leq p_Z$ of the region.  The thresholds for this $5$ in $16$ code are given in level 5 of Tab. \ref{tab:5-64}.  Therefore the entire region around $(p,0,p)$ is correctable with a combination of these two codes.  It follows that we can correct all Pauli noise below the hashing bound, disproving the conjecture made in \cite{SS}. It is still an open question whether this result applies to all non Pauli noise. 

We then find the optimal $n_1$ qubit bit flip code for a given noise by optimizing over all lengths $n_1 \leq 150$. The optimal value of $n_1$ is always at least $5$, and is approximately $n_1 \approx \frac{2}{p_1}$. For small $p_1$, Prop. \ref{prop:infinitenoise} gives a threshold of $(p_X,p_Y,p_Z) \approx (\frac{1}{4}\epsilon^2,\frac{1}{2}-\epsilon, \frac{1}{4}\epsilon^2)$, which is very similar to that predicted by Prop. \ref{prop:infinitenoise} in the case of small $q_Z$ for independent noise.

\begin{figure*}
\includegraphics[width=0.77\textwidth]{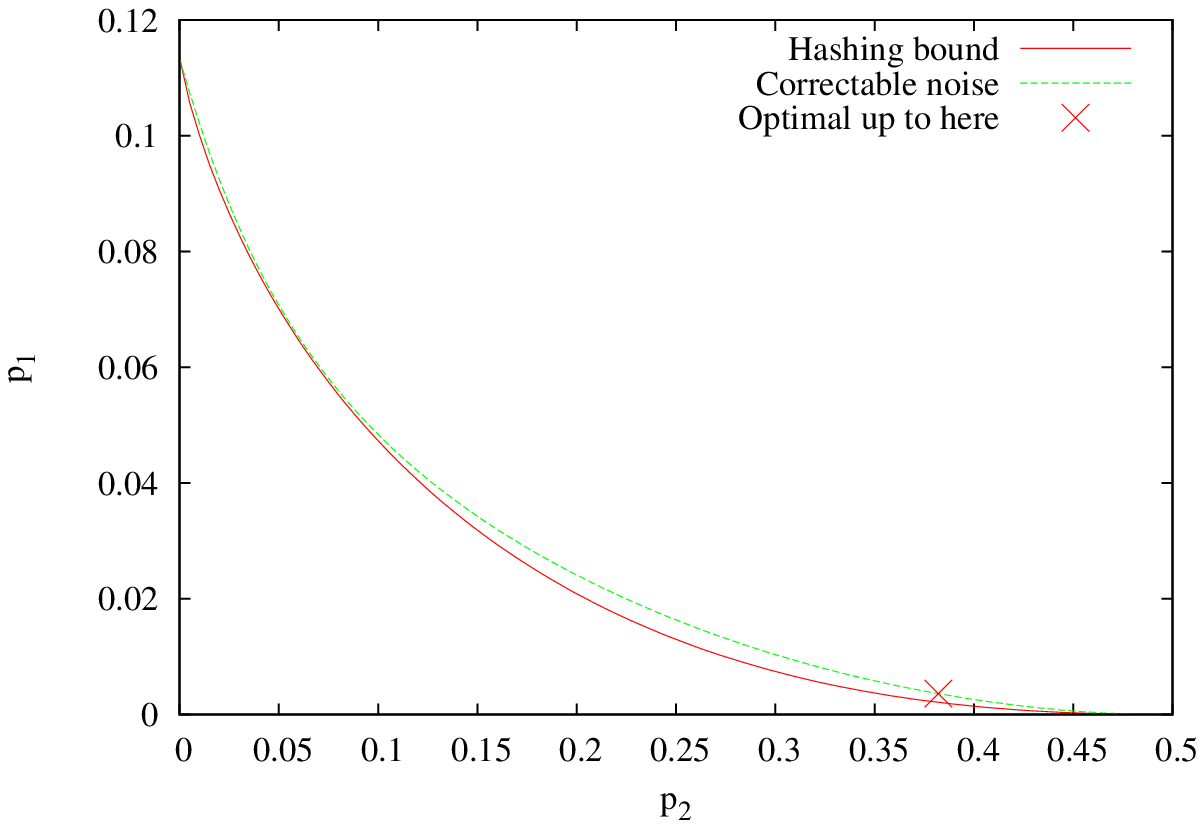}
\caption{\label{fig:nind}
\capt Threshold of the non-zero capacity for two Pauli noise $(p_X,p_Y,p_Z)=(p_1,p_2,p_1)$. For each type of noise, we use the optimal $n_1$ qubit bit flip code with $n_1 \leq 150$. To the right of the "optimal up to here" point, a code with $n_1 > 150$ would be optimal. These threshold values are determined exactly.}
\end{figure*}

In Fig. \ref{fig:nind}, we show the threshold values of correctable two-Pauli noise as a function of $p_1$ and $p_2$ for $n_1$ qubit bit flip codes with $n_1 \leq 150$.  

\begin{table}
\caption{\label{tab:bounds} 
Bounds on non-zero capacity for different types of $(p_X, p_Y, p_Z)$ noise. The hashing bound is obtained when the entropy is $1$. The bit flip code is the threshold from applying the optimal bit flip code of length at least two. Bit/phase is the threshold from applying the optimal bit flip and then phase flip code combination. The lower bound given is the result of taking the previous result and concatenating with the $[[5,1,3]]$ code repeatedly, as described in this paper. These are compared to the upper bound.}
\begin{ruledtabular}
\begin{tabular}{clll}
             & $(p,p,p)$  & $(p-p^2,p^2,p-p^2)$ & $(p,0,p)$ \cr
\hline
Hashing   & 6.30965616\%   &  11.00278644\%  &  11.35460976\%\cr 
Bit flip  & 6.34520294\%   &  11.21074102\%  &  11.33392680\% \cr
Bit/phase & 6.37338273\%   &  11.27458434\%  &  11.39425214\% \cr
Lower     & 6.3767(5)\%    &  11.280(6)\%    &  11.400(3)\% \cr
Upper     & 8.3333\%       &  14.6447\%      &  16.6667\% \cr
\end{tabular}
\end{ruledtabular}
\end{table}

\begin{table}
\caption{\label{tab:entropy} 
Shannon entropies of the channels at the thresholds in Tab. \ref{tab:bounds}}
\begin{ruledtabular}
\begin{tabular}{clll}
             & $(p,p,p)$  & $(p-p^2,p^2,p-p^2)$ & $(p,0,p)$ \cr
\hline
Hashing   & 1              &  1              &  1 \cr 
Bit flip  & 1.00392304     &  1.01248000     &  0.99885469 \cr
Bit/phase & 1.00702529     &  1.01628620     &  1.00219124 \cr
Lower     & 1.00739(6)     &  1.0166(4)      &  1.0025(3)\cr
Upper     & 1.20752        &  1.20175        &  1.25163 \cr
\end{tabular}
\end{ruledtabular}
\end{table}

\section{Infinite length bit flip code}
\label{sec:infinite}
It can also be interesting to determine the thresholds for a bit flip code of infinite length. We find a threshold in terms of arbitrary $(p_X, p_Y, p_Z)$ values below.

\begin{proposition}
\label{prop:infinitenoise}
Suppose a bit flip code with an infinite number of qubits $n$ is used to correct independent noise on each qubit. There is a critical threshold of 
\begin{equation*}
2 (q_X(1-q_X))^{\frac{3}{2}} < (p_X-p_Y)^2 (1-q_X)+(p_I-p_Z)^2 q_X.
\end{equation*}
If $2 (q_X(1-q_X))^{\frac{3}{2}}$ is below this, the logical entropy will be less than $1$, and if $2 (q_X(1-q_X))^{\frac{3}{2}}$ is above this, the logical entropy will be greater than $1$. By concatenating by the optimal $n_2$ qubit phase flip code, these will go to $0$ (perfectly correctable noise) and $2$ (completely depolarizing noise), respectively.
\end{proposition}
\begin{proof}
From Stirling's approximation, 
\begin{equation*}
\binom{n}{fn} \approx \frac{1}{\sqrt{2 \pi}} \frac{n^{n + \frac{1}{2}}}{(fn)^{fn+\frac{1}{2}} (n-fn)^{n-fn+\frac{1}{2}}},
\end{equation*}
we have the formula
\begin{equation*}
\lim_{n \rightarrow \infty} \sqrt[n]{\binom{n}{fn}} = \frac{1}{f^f (1-f)^{1-f}} = 2^{H(f,1-f)}
\end{equation*}
that we shall use.

As $n \rightarrow \infty$, the dominant contribution to the $X$ entropy $X_e$ derives from $l_I(k)$ and $l_X(k)$ for $k$ near $\frac{n}{2}$. Therefore 
\begin{align}
\label{eq:Xent}
\lim_{n \rightarrow \infty} \sqrt[n]X_e = \lim_{n \rightarrow \infty} \sqrt[n]{l_I(\frac{n}{2})} \nonumber \\
= \lim_{n \rightarrow \infty} \sqrt[n]{a_{\frac{n}{2}}} = 2 \sqrt{q_X (1-q_X)}.
\end{align}
The $Z$ entropy is dominated by the sum of the entropy contributions from $l_I(k)$ and $I_Z(k)$, which is given by
\begin{equation*}
Z_e = \sum_k a_k (H(\frac{1}{2}+\frac{b_k}{2 a_k} , \frac{1}{2} - \frac{b_k}{2 a_k}).
\end{equation*}
Up to second order, $H(\frac{1}{2}+\epsilon, \frac{1}{2} - \epsilon) \approx 1 - c \epsilon^2$ where $c = 2 \log_2 e$; therefore
\begin{equation*}
Z_e \approx \sum_k a_k (1-c (\frac{b_k}{2 a_k})^2) \approx 1 - \frac{c}{4} \sum_k \binom{n}{k} g^k h^{n-k},
\end{equation*}
where $g = \frac{(p_X - p_Y)^2}{q_X}$ and $h = \frac{(pI-p_Z)^2}{1-q_X}$. Letting $k = fn$, 
\begin{equation*}
\lim_{n \rightarrow \infty} \sqrt[n]{1-Z_e} = \max_f (\frac{g}{f})^f (\frac{h}{1-f})^{1-f},
\end{equation*}
which is maximum for $f = \frac{g}{g+h}$; therefore
\begin{equation*}
\lim_{n \rightarrow \infty} \sqrt[n]{1-Z_e}  = g + h = \frac{(p_X - p_Y)^2}{q_X} + \frac{(p_I-p_Z)^2}{1-q_X}.
\end{equation*}
To obtain the inequality in the proposition, we use this and $\lim_{n \rightarrow \infty} \sqrt[n]{X_e}$ from Eq. \ref{eq:Xent}, and note that $\lim_{n \rightarrow \infty} X_e + Z_e < 1$ if and only if 
\begin{equation*}
\lim_{n \rightarrow \infty} \sqrt[n]{X_e} < \lim_{n \rightarrow \infty} \sqrt[n]{1-Z_e}.
\end{equation*}
Hence
\begin{equation*}
2 (q_X(1-q_X))^{\frac{1}{2}} < (p_X-p_Y)^2 +(p_I-p_Z)^2,
\end{equation*}
and multiplying by $q_X(1-q_X)$ leads to the desired result.

\end{proof}
The entropy is $0$ when there is only one type of Pauli error, and reaches its maximal value of $2$ for the completely depolarizing noise $(p_X, p_Y, p_Z) = (\frac{1}{4}, \frac{1}{4}, \frac{1}{4})$.

\section{Summary of bounds}
We now compare the lower bounds derived above to some known upper bounds on non-zero capacity that are given by \cite{Cerf}:
\begin{equation}
p_X + p_Y + p_Z + \sqrt{p_X p_Y} + \sqrt{p_Y p_Z} + \sqrt{p_Z p_X} < \frac{1}{2}.
\end{equation}
We can use this expression to obtain upper bounds on the types of noise studied in this work. The corresponding upper bounds are given in  Tab. \ref{tab:bounds}. As stated previously, the hashing bound is when the entropy is $1$ without applying a code. The bit flip code rates are the maximum attainable with a bit flip code. The next row is the maximum obtainable with a bit flip code concatenated with a phase flip code. The best known lower bounds on the thresholds are those described earlier in this work. They were found by repeatedly concatenating the bit flip in phase flip code ($n_1$ in $n_2$) with the $[[5,1,3]]$ code. The Shannon entropies of the channels at the threshold values from Tab. \ref{tab:bounds} are given in Tab. \ref{tab:entropy}. Note that the channel entropy depends on the type of noise. In particular, the best known improvement over the hashing bound is $6$ times as high for the $(p-p^2,p^2,p-p^2)$ type noise as for the $(p,0,p)$ type noise. 

In Tab. \ref{tab:bounds}, we see that the pure bit flip codes can correct above the hashing bound except near $(p,0,p)$. Further improvement is found by concatenating the bit flip code with a phase flip code; this always corrects up to at least the hashing bound for any Pauli noise.  Repeated concatenation with the $[[5,1,3]]$ code gives some minor additional improvement. Clearly, there is a large gap between the best known lower and upper bounds. We conjecture that the ultimate thresholds on the non-zero capacity are much closer to the lower bounds than the upper bounds in Tab. \ref{tab:bounds}. However, we expect that the new lower bounds found in this paper will not be the ultimate threshold values.

\section{Summary and Discussion}
In this paper we have established new lower bounds for the non-zero capacity of noise quantum channels, using an entropic approach to calculation of thresholds of correctable noise.  Our analysis consisted of two steps: first, performing exact calculations to find the thresholds below which the $n_1$ in $n_2$ code results in a logical entropy of less than $1$. We then further concatenate the $n_1$ in $n_2$ code a number of times $j$ with the $[[5,1,3]]$ codes and use Monte Carlo simulation to evaluate the corresponding threshold values.  We applied the procedure to various kinds of Pauli noise and found that this double concatenation yields improved threshold values over previously existing codes in all cases.  Our results show that Pauli noise is correctable up to the hashing bound, disproving a recent conjecture that there exists some uncorrectable Pauli noise below the hashing bound~\cite{SS}.

The doubly concatenated codes presented here result in one logical qubit encoded in $n_1 n_2 5^j$ physical qubits.  They can then be further concatenated with random codes, as in \cite{DiVincenzo98,SS}, to generate a non-zero capacity code.  
Unfortunately, since extrapolation of the Monte Carlo simulation results to zero error introduces some uncertainty by virtue of their stochastic nature, these doubly concatenated codes do not provide a constructive route to non-zero capacity codes in the current analysis.  The results are nevertheless strongly suggestive that a constructive procedure based on these or related codes might be possible.  The doubly-concatenated encoding analyzed here thus provides a basis for further efforts towards the  challenge of finding constructive encodings for finite capacity quantum channels.

\begin{acknowledgments}
We thank the NSF for financial support under ITR Grant No. EIA-0205641 and Y. Ouyang for a useful discussion. 
\end{acknowledgments}


\begin{thebibliography}{10}

\bibitem{MoreChans}
J.~Fern.
\newblock Correctable noise of quantum error correcting codes under adaptive
  concatenation.
\newblock {\em Phys. Rev. A}, 77:010301 (R), 2008.
\newblock quant-ph/0703258.

\bibitem{Shannon}
C.~E. Shannon.
\newblock A mathematical theory of communication.
\newblock {\em Bell Syst. Tech. J.}, 27:379, 1948.

\bibitem{SS}
G.~Smith and J.~Smolin.
\newblock Degenerate quantum codes for pauli channels.
\newblock {\em Phys. Rev. Lett.}, 98(3):030501, 2007.
\newblock quant-ph/0604107.

\bibitem{Lloyd}
Seth Lloyd.
\newblock Capacity of the noisy quantum channel.
\newblock {\em Physical Review A}, 55:1613--1622, 1997.

\bibitem{Devetak2003}
I.~Devetak.
\newblock The private classical capacity and quantum capacity of a quantum
  channel.
\newblock {\em IEEE Trans. Information Theory}, 51(1):44--55, 2005.
\newblock quant-ph/0304127.

\bibitem{ShorMSRI}
P.~W. Shor.
\newblock The quantum channel capacity and coherent information.
\newblock MSRI Workshop in Quantum Computation 2002.
  \url{http://www.msri.org/publications/ln/msri/2002/quantumcrypto/shor/1/}.

\bibitem{Schumacher}
B.~Schumacher and M.~A. Nielsen.
\newblock Quantum data processing and error correction.
\newblock {\em Phys. Rev. A}, 54:2629--2635, 1996.
\newblock quant-ph/9604022.

\bibitem{BDSW}
C.~Bennett, D.~DiVincenzo, J.~Smolin, and W.~Wootters.
\newblock Mixed state entanglement and quantum error correction.
\newblock {\em Phys. Rev. A}, 54:3824, 1996.
\newblock quant-ph/9604024.

\bibitem{ShorSmolin}
P.~W. Shor and J.~Smolin.
\newblock Quantum error-correcting codes need not completely reveal the error
  syndrome.
\newblock quant-ph/9604006.

\bibitem{DiVincenzo98}
D.~P. DiVincenzo, P.~W. Shor, and J.~Smolin.
\newblock Quantum channel capacity of very noisy channels.
\newblock {\em Phys. Rev. A}, 57(2):830, 1998.
\newblock quant-ph/9706061.

\bibitem{LMPZ}
R.~Laflamme, C.~Miquel, J.P. Paz, and W.H. Zurek.
\newblock Perfect quantum error correction code.
\newblock {\em Phys. Rev. Lett.}, 77:198--201, 1996.
\newblock quant-ph/9602019.

\bibitem{Cerf}
N.~Cerf.
\newblock Asymmetric quantum cloning in any dimension.
\newblock {\em J. Mod. Opt.}, 47:187--209, 2000.
\newblock quant-ph/9805024.

\end{thebibliography}
\end{document}